\documentclass{article}
\usepackage[final]{neurips_2019}

\usepackage[utf8]{inputenc} 
\usepackage[T1]{fontenc} 
\usepackage{hyperref} 
\usepackage{url} 
\usepackage{graphicx}
\usepackage{booktabs} 
\usepackage{amsfonts} 
\usepackage{amsmath,amssymb,amsthm}
\usepackage{nicefrac} 
\usepackage{microtype} 

\graphicspath{
}

\title{Transfer Learning of {fMRI} Dynamics}

\author{%
 Usman Mahmood\thanks{Equal contribution} \\
 Georgia State University\\
 \texttt{umahmood1@student.gsu.edu}
 \And
 Md Mahfuzur Rahman\footnotemark[1]\\
 Georgia State University \\
 \texttt{mrahman21@student.gsu.edu}
 \And
 Alex Fedorov\footnotemark[1] \\
 Georgia Institute of Technology \\
 \texttt{afedorov@gatech.edu}
 \AND
 Zening Fu \\
 TReNDS Center\\
 Georgia State University\\
 \texttt{zfu@gsu.edu}
 \And
 Sergey Plis\\
 TReNDS Center\\
 Georgia State University\\
 \texttt{s.m.plis@gmail.com }
}

\begin{document}

\maketitle

\begin{abstract}
 As a mental disorder progresses, it may affect brain structure, but brain function expressed in brain dynamics is affected much earlier. Capturing the moment when brain dynamics express the disorder is crucial for early diagnosis. The traditional approach to this problem via training classifiers either proceeds from handcrafted features or requires large datasets to combat the $m>>n$ problem when a high dimensional fMRI volume only has a single label that carries learning signal. Large datasets may not be available for a study of each disorder, or rare disorder types or sub-populations may not warrant for them. In this paper, we demonstrate a self-supervised pre-training method that enables us to pre-train directly on fMRI dynamics of healthy control subjects and transfer the learning to much smaller datasets of schizophrenia. Not only we enable classification of disorder directly based on fMRI dynamics in small data but also significantly speed up the learning when possible. This is encouraging evidence of informative transfer learning across datasets and diagnostic categories.
 
\end{abstract}

\section{Introduction}
Mental disorders manifest in behavior that is driven by disruptions in brain dynamics. Functional MRI captures the nuances of spatiotemporal dynamics that could potentially provide clues to the causes of mental disorders and enable early diagnosis. However, the obtained data for a single subject is of high dimensionality $m$, and to be useful for learning, and statistical analysis, one needs to collect datasets with a large number of subjects $n$. Yet, for any disorder, demographics, or other types of conditions, a single study is rarely able to amass datasets large enough to go out of the $m>>n$ mode. Traditionally this is approached by handcrafting features~\cite{Khazaee2016} of a much smaller dimension, effectively reducing $m$ via dimensionality reduction. Often, the dynamics of brain function in these representations vanishes into proxy features such as correlation matrices of functional network connectivity (FNC)~\citep{yan2017discriminating}. Efforts that pull together data from various studies and increase $n$ do exist, but it is difficult to generalize to the study of smaller and more specific disease populations that cannot be shared to become a part of these pools or are too different from the data in them.

Our goal is to enable the direct study of brain dynamics in smaller datasets to, in turn, allow an analysis of brain function. In this paper, we show how one can achieve significant improvement in classification directly from dynamical data on small datasets by taking advantage of publicly available large but unrelated datasets. We demonstrate that it is possible to train a model in a self-supervised manner on the dynamics of healthy control subjects from the Human Connectome Project (HCP)~\citep{van2013wu} and apply that pre-trained encoder to a completely different data collected across multiple sites from healthy controls and schizophrenia subjects.

\section{Related Work}

Recent advances in unsupervised learning using self-supervised methods by estimating and maximizing mutual information reduced the gap between supervised and unsupervised learning~\citep{oord2018representation, hjelm2018learning, bachman2019learning}. Such success has already influenced neuroimaging in the case of structural MRI~\citep{fedorov2019prediction} and even reinforcement learning~\citep{anand2019unsupervised}. 

Prior works in brain imaging have been based on unsupervised methods such as linear ICA~\citep{calhoun2001method} and HMM framework~\cite{eavani2013unsupervised}. Some other nonlinear approaches were also proposed to capture the dynamics as using RBMs~\citep{hjelm2014restricted} and RNN modification of ICA~\citep{hjelm2018spatio}.

Also, in most cases, researchers in brain imaging are dealing with small datasets. In this case, transfer learning~\citep{mensch2017learning, 10.3389/fnins.2018.00491, thomas2019deep} might be a way to improve results and in some cases, to enable learning from data otherwise too small for any results. Another idea to improve performance might be considered by a data generating approach~\citep{ulloa2018improving}.

\section{Method Description}

For self-supervised pre-training, we are using spatio-temporal objective ST-DIM~\citep{anand2019unsupervised} to maximize predictability between current latent state and future spatial state and between consecutive spatial states. For the lower bound of mutual information, we are using InfoNCE~\citep{oord2018representation} estimator. Compare to other available estimators, InfoNCE shows better performance~\citep{hjelm2018learning, bachman2019learning} in case of a greater number of negative samples that are readily available in case of time series data. 

Let $\{(u_t, v_s): 1 \le t, s \le N, t \ne s\}$ be a dataset of pairs of values at time point $t$ and $s$ sampled from sequence with length $N$. 
A pair $(u_t, v_s)^+$ is called positive if $s = t + 1$ and $(y_t, v_s)^-$ --- negative if $s \ne t + 1$. A positive pair models the joint and a negative --- marginal distributions. Eventually, the InfoNCE estimator is defined as:
\begin{equation}
 \mathcal{I}_f\Big(\{(u_t, v_{t+1})^+\}^N_{t=1}\Big) = \sum^N_{t=1} \log \frac{\exp f((u_t, v_{t+1})^+)}{\sum^N_{s=1} \exp f((u_t, v_s)^-)},
\end{equation}
where $f$ is a critic function~\cite{tschannen2019mutual}. Specifically, we are using separable critic $f(u_t,v_s) = \phi(u_t)^\intercal \psi(v_s)$, where $\phi$ and $\psi$ are some embedding function parametrized by neural networks. Such embedding functions are used to calculate value of a critic function in same dimensional space from two dimesionally different inputs. Critic learns an embedding function such that it assigns higher values for positive pairs compare to negative pairs: $f((u_t, v_{t+1})^+) \gg f((u_t, v_s)^-)$.

We define a latent state as an output $z_t$ of encoder $E$ and a spatial state $c^l_t$ as the output of $l$th layer of the encoder for input $x_t$ at time point $t$. To optimize the objective between current latent state and future spatial state the critic function for input pair $(x_t, x_s)$ is $f_{LS} = \phi(z_t)^\intercal \psi(c^l_s)$ and for consecutive spatial states --- $f_{SS} = \psi(c^l_t)^\intercal \psi(c^l_s)$. Finally, the loss is the sum of the InfoNCE with $f_{LS}$ and InfoNCE with $f_{SS}$ as $L = {I}_{f_{LS}} + {I}_{f_{SS}}$.

\section{Experiments}

\subsection{Datasets}

\subsubsection{Simulation}

To simulate the data we generate multiple $10$-node graphs with $10 \times 10$ stable transition matrices. Using these we generated multivariate time series with autoregressive (VAR) and structural vector autoregressive (SVAR) models~\citep{lutkepohl2005new}. 

First, we generate $50$ VAR times series with size $10 \times 20000$. Then we split our dataset to $50\times10\times14000$ samples for training, $50\times10\times4000$ ---for validation and $50\times10\times2000$ --- for testing. Using these samples we pre-train an encoder and evaluate based on its ability to identify consecutive $10 \times 20$ windows sampled from whole time series.

In the final downstream task we classify the whole time-series whether it is generated by VAR or SVAR (undersampled VAR at rate 2). We create $400$ graphs with corresponding stable transition matrices and generate $2000 \times 10 \times 4000$ samples ($5$ for each) and split as $1600\times 10 \times 4000$ for training, $200\times 10 \times 4000$ for validation and $200\times 10 \times 4000$ for hold-out test. Here we also use $10 \times 20$ windows as a single time-point input.

\subsubsection{Real data}

Two independent datasets were used in the current study. 
The first dataset is a Schizophrenia dataset, which is from Function Biomedical Informatics Research Network (FBIRN)~\citep{keator2016function}\footnote[1]{These data were downloaded from the Function BIRN Data Repository, Project Accession Number 2007-BDR-6UHZ1}), and the second dataset is a healthy subject dataset, which is from the 1200 Subject release of Human Connectome Project (HCP)~\citep{van2013wu}.

The FBIRN dataset was pre-processed through SPM12~\citep{penny2011statistical} based on the MATLAB 2016b environment. 
The slice-timing was first performed on the data, and then subject head motions were corrected by the realignment procedure. 
After that, the data was warped to MNI space using EPI template and resampled to $3$ mm$^3$ voxels. 
Finally, the data were smoothed with a 6mm FWHM Gaussian kernel. 
The FBIRN dataset consists of $311$ subjects, including $151$ SZ patients and $160$ healthy controls.

The resting-state fMRI HCP data comes pre-processed by the following pipeline~\citep{glasser2013minimal}. 
It includes removing of spatial distortions, compensation of the subject motion, reduction of the bias field, normalization with a global mean, and final brain masking. 
The pre-processed HCP data were then warped to the MNI space using the EPI template and resampled to the $3$ mm$^3$ voxels using the same bounding box, to guarantee HCP and FBIRN datasets have the same spatial resolution and dimensions. HCP consists only of healthy controls.

For each dataset, $53$ intrinsic connectivity networks (ICNs) were extracted using the pipeline described at~\citep{Du19008631}. 
These $53$ ICNs are supposed to be non-noise components providing meaningful functional network information and thus were used in training.

\subsection{Training}

Encoder for simulation experiment consist of $4$ $1$D convolutional layers with output features $(32,64,128,64)$, kernel sizes $(4,4,3,2)$ and stride --- $1$, following by ReLU~\cite{glorot2011deep} after each layer followed by Linear layer with $256$ units. For real data --- $3$ $1$D convolutional layers with $(64, 128, 200)$, $(4, 4, 3)$ and $1$ respectively, followed by linear layer with $256$ units. Then for all possible pairs in the batch we took flattened features after $3$rd convolutional layer $c^3$ and features from last layer $z$. We embedded them using $\psi$ for $c^3$ and $\psi$ for $z$ to $128$ dimensional vector 
to compute the score of a critic function $f_{LS}$ or $f_{SS}$. Using these scores we computed the loss. The neural networks trained using Adam optimizer~\citep{kingma2014adam}. The weights were initialized using Xavier~\citep{glorot2010understanding}.

For simulation experiment, first, we train our encoder to learn on $10 \times 20$ windows from the VAR time series using InfoNCE based loss, and secondly, we train a supervised classifier based on windows. This window-based classification provides promising results (accuracy $~60\%$). However, in solving similar real problems, we are more interested in subjects, i.e., entire time series, rather than a single-window for classification. Hence, we perform classification based on the whole time-series. In this setting, the entire time-series is encoded as a sequence of representations and fed through a biLSTM classifier. Two additional linear layers with $200$ hidden units on top of the last hidden state of the biLSTM are used to map the representation to classification scores. 

For the real data case, similar to simulations, we successfully train (accuracy $~87-90 \%$) our encoder on consecutive windows of fMRI from HCP healthy subjects. Then each computed feature for each window of the whole fMRI sequence used to train biLSTM classifier on fBIRN dataset. The biLSTM classifies SZ and HC subjects. Overall each fMRI of the subject consists of a series of $13$ overlapping by half windows by $53$ components by $20$ time points.

\begin{figure}[ht]
\begin{minipage}{.45\linewidth}
 \includegraphics[width=0.8\linewidth]{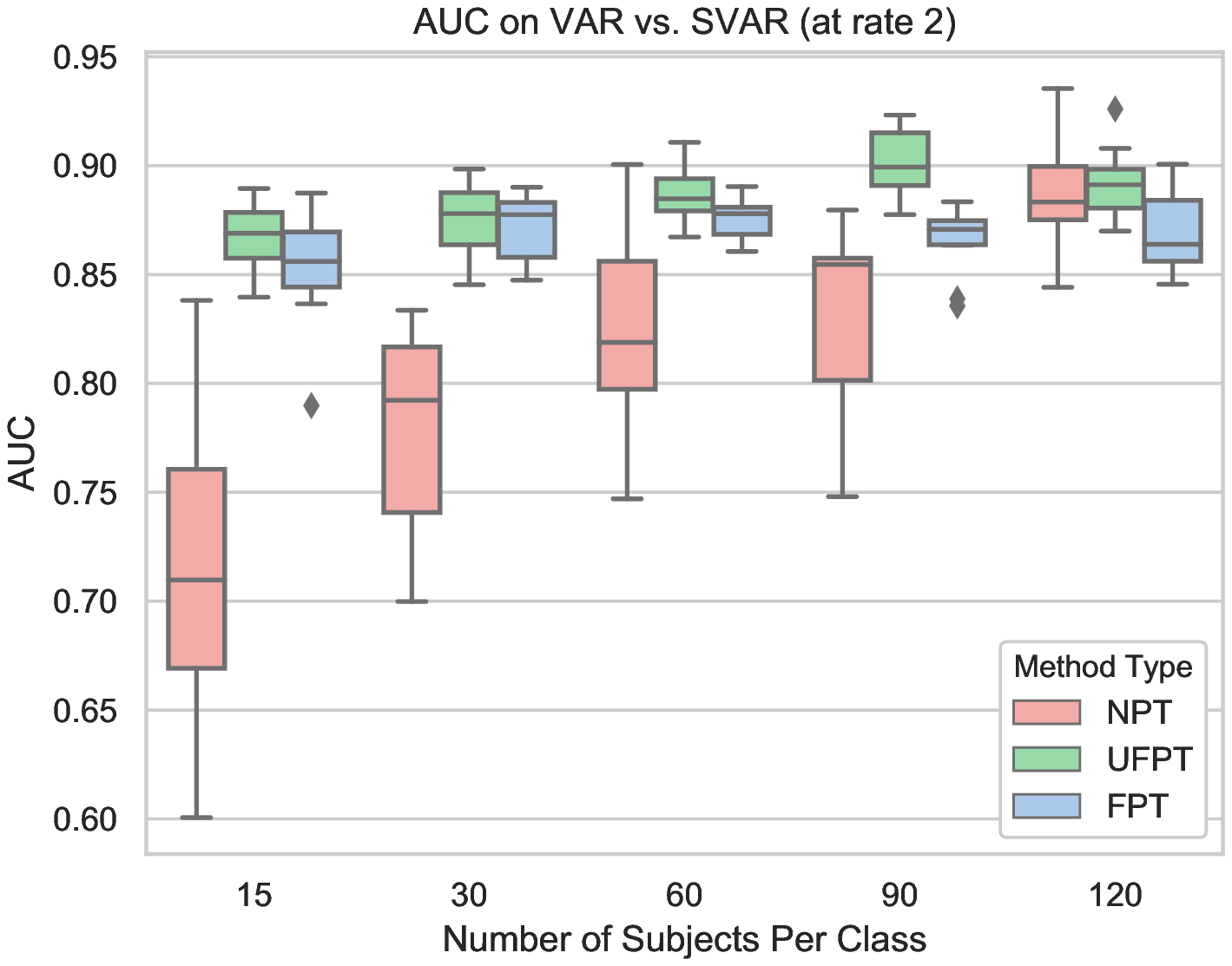}
 \centering
 \caption{VAR vs. SVAR time-series classification accuracy of synthetic data.}
 \label{fig:synth_test}
 \vspace{-1mm}
\end{minipage}%
\hfill
\begin{minipage}{.45\linewidth}
 \includegraphics[width=0.8\linewidth]{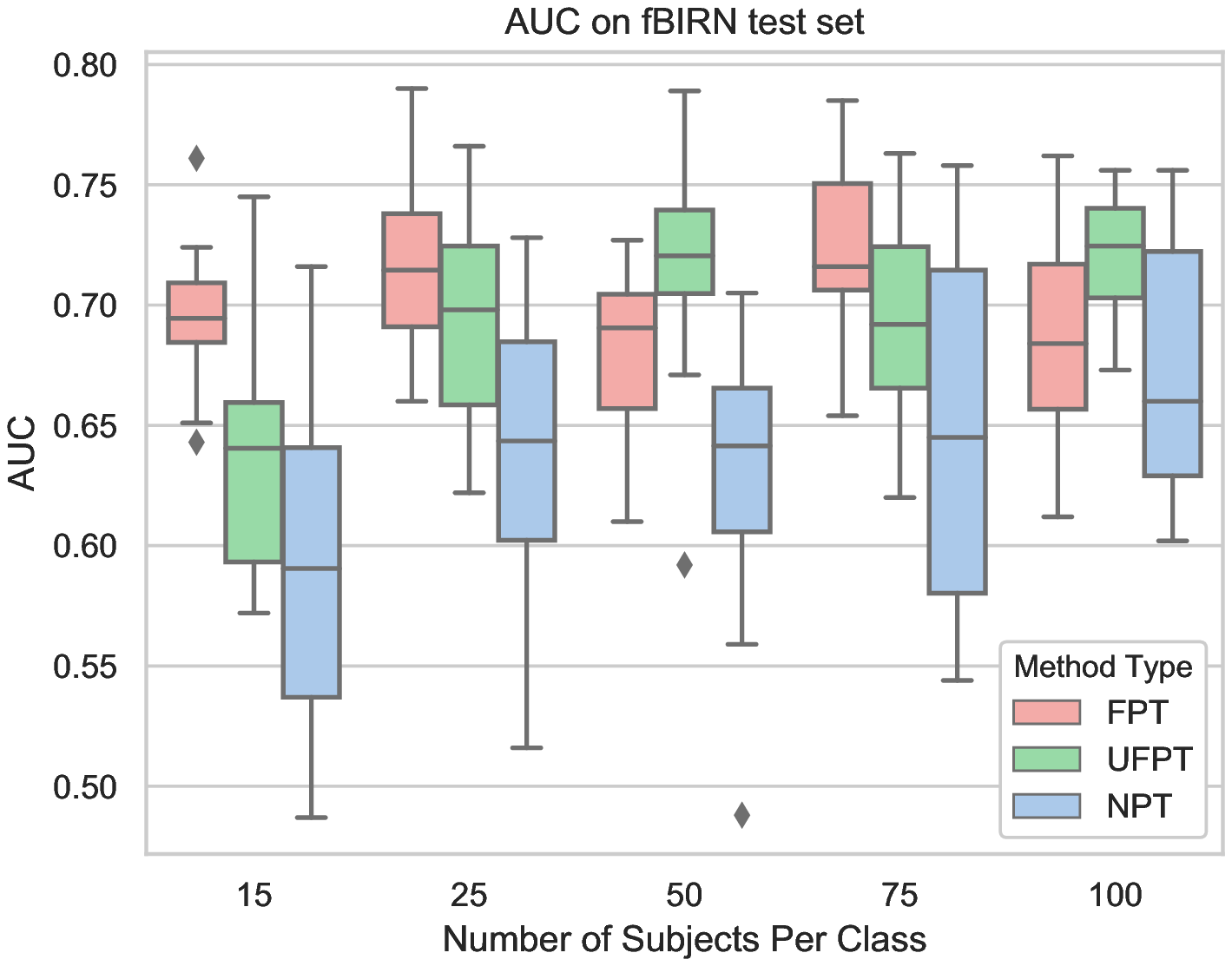}
 \centering
 \caption{Classification results on the progressively larger datasets.}
 \label{fig:real_test}
 \vspace{-1mm}
 \end{minipage}
\end{figure}

\subsection{Results}

Here we compare an end-to-end supervised model without pre-training (NPT), with frozen layers of the pre-trained encoder (FPT), and with unfrozen layers of the pre-trained encoder (UFPT).

In the simulation study, we observe that the pre-trained model can easily be fine-tuned only with a small amount of downstream data. Our model can classify a randomly chosen time-series as a sample of VAR or SVAR~(Figure~\ref{fig:synth_test}). Note, with very few training samples, models based on the pre-trained encoder outperform supervised models. However, as the number of samples grows, the accuracy achieved with or without pre-training levels out.

As we can see from Figure~\ref{fig:real_test}, the real data results substantiate the insights achieved in a simulation study. The test dataset consists of $64$ subjects that are held out from training and validation processes and are the same for all tests in the plot. Training data was
randomly resampled ten times from the available data pool. To put it another way, self-supervised transferable pre-training always helps when we have very few samples offering higher AUC.

\section{Conclusions and Future Work}

As we have demonstrated, self-supervised pre-training of a spatiotemporal encoder on fMRI of healthy subjects provides benefits that transfer across datasets, collection sites, and to schizophrenia disease classification. Learning dynamics of fMRI helps to improve classification results for schizophrenia on small datasets, that otherwise do not provide reliable generalizations. Although the utility of this result is highly promising by itself, we conjecture that direct application to spatiotemporal data will warrant benefits beyond improved classification accuracy in the future work. Working with ICA components is smaller and thus easier to handle space that exhibits all dynamics of the signal. In the future, we will move beyond ICA pre-processing and work with fMRI data directly. We expect model introspection to yield insight into the spatio-temporal biomarkers of schizophrenia. In future work, we will test the same analogously pre-trained encoder on datasets with various other mental disorders such as MCI and bipolar. We are optimistic about the outcome because the proposed pre-training is oblivious to the downstream use and is done in a manner quite different from the classifier's work. It may indeed be learning crucial information about dynamics that might contain important clues into the nature of mental disorders.

\section{Acknowledgement}
This study is supported by NIH grant R01EB020407.

Data were provided [in part] by the Human Connectome Project, WU-Minn Consortium (Principal Investigators: David Van Essen and Kamil Ugurbil; 1U54MH091657) funded by the 16 NIH Institutes and Centers that support the NIH Blueprint for Neuroscience Research; and by the McDonnell Center for Systems Neuroscience at Washington University.
Additional data used in this study were downloaded from the Function
BIRN Data Repository (http://fbirnbdr.birncommunity.org:8080/BDR/),
supported by grants to the Function BIRN (U24-RR021992) Testbed funded by the
National Center for Research Resources at the National Institutes of Health, U.S.A.

\bibliographystyle{plainnat}
\bibliography{references}

\end{document}